\documentclass[showpacs,preprint,aps,superscriptaddress,groupedaddress,nofootinbib]{revtex4-1}

\usepackage{graphicx}
\usepackage{dcolumn}   
\usepackage{bm}       
\usepackage{amssymb}
\usepackage{amsmath}
\newcommand{\del}{\partial}

\newcommand{\Ss}{\sqrt{\sigma}}
\newcommand{\goes}{\rightarrow}

\newcommand{\A}{\alpha}
\newcommand{\B}{\beta}
\newcommand{\G}{\gamma}
\newcommand{\T}{\theta}
\newcommand{\p}{\phi}
\newcommand{\e}{\eta}
\begin{document}
	\title{Magnetic monopole as a spacetime defect}
	
	\author{Suvikranth Gera}
	\email{suvikranthg@iitkgp.ac.in}
	\author{Sandipan Sengupta}
	\email{sandipan@phy.iitkgp.ac.in, sandipan422@gmail.com}
	\affiliation{Department of Physics, Indian Institute of Technology Kharagpur, Kharagpur-721302, India}
		
	\begin{abstract}

		%We show that a magnetic charge in curved spacetime could be an artefact of a vacuum phase with zero metric determinant at a distance. This phase is characterized by a solution of the first order field equations with nontrivial torsion. The (apparent) monopole charge here has a topological origin, given precisely by a lower dimensional counterpart of the Nieh-Yan invariant. In this geometric realization, the monopole core remains hidden from the observer  living in the invertible metric phase, thus precluding its direct detection.
		%%
		
		We propose that the only possible realization of a magnetic pole is emergent, it being an artefact
		of a torsion defect in a curved spacetime. This special phase is characterized by a class of degenerate metric spacetime solutions of first order gravity in vacuum. The (apparent) magnetic charge  is shown to have a topological origin, given by a lower dimensional counterpart of the Nieh-Yan invariant. At the invertible metric phase at a distance, this topological charge gets reflected as the (magnetic) Reissner-Nordstrom charge to an asymptotic observer, even though the defect itself remains hidden.
	\end{abstract}	

	\maketitle

\section{Introduction}
	Dirac's formulation of singular magnetic monopoles stands out as a remarkably elegant yet unconfirmed explanation of the quantization of electric charge \cite{dirac,dirac1}. As 't hooft and Polyakov showed later, monopole fields without string singularities could emerge naturally in non-abelian gauge theories as  solitonic solutions of finite energy and size and a definite magnetic charge \cite{thooft,polyakov,prasad,bogo}. Their presence also appears to be fundamental in the formulation of grand unified theories.  However, the lack of any experimental encouragement till now makes one wonder whether monopoles really do exist in nature, at least in the forms they have been theorized till now.

	Here we work up the slightly more radical idea that the magnetic charge in a curved spacetime could be a purely gravitational (emergent) effect. The apparent charge is seeded by a special phase of first order gravity theory, where the associated classical spacetime solution necessarily has a noninvertible four-metric \cite{tseytlin,kaul1}. To emphasize, this description does not involve any matter-coupling such as non-abelian gauge fields. 
	
	Based on the field equations of first-order vacuum gravity, we provide an emergent definition for the magnetic
field and the associated magnetic charge. The noninvertible solution necessarily exhibits a nontrivial spatial contortion, whose form is constrained by eliminating the Dirac-string singularities in the spacetime geometry. The emergent three-geometry where the apparent charge lives is regular, as the evaluation of the characteristic curvature scalars explicitly shows.
	
	The degenerate core of the apparent monopole cannot be observed directly. This is because it cannot be reached along any timelike radial trajectory (geodesic) by an observer outside, living in a standard Einsteinian geometry. Locally, this part of the spacetime is equivalent to the (extremal) magnetic  Reissner-Nordstrom solution almost everywhere. The full spacetime is spherically symmetric and smooth.  
	
Since there is no genuine matter source, the apparent magnetic charge
must be a reflection of a topological charge of the noninvertible metric
solution in vacuum. We demonstrate that this is indeed the case, the
relevant topological invariant being a lower dimensional counterpart of
the Nieh-Yan number. In other words, what is manifested as a magnetic
Reissner-Nordstrom charge in curved spacetime away from the vacuum
phase has a purely topological interpretation within this formulation. The
implications of such a connection could be radical.	
	
	This construction of an emergent monopole as a spacetime defect in gravity theory provides a potential theoretical justification as to why a free magnetic pole is not observed in nature. In this sense, this configuration supercedes the nonsingular 't hooft-Polyakov monopole built upon genuine matter source, which should in principle be detected experimentally.

\section{Magnetic charge in first order gravity}	
In order to define a `static' solution in gravity theory, an appropriate setting is to consider a spacetime configuration where the metric has a degenerate eigenvalue along the timelike direction, which naturally precludes  any temporal dynamics. In general, such three-geometries could emerge as solutions to the field equations of first-order gravity \cite{tseytlin,kaul1,kaul,sandipan,sengupta,bengtsson,gera}, where the invertibility of tetrads is not assumed or required apriori\footnote{Tetrad compatibility condition would be assumed throughout.}.

		Let us consider a spacetime geometry given by the following four-metric:
	\begin{eqnarray}\label{metric}
	ds^2&=& 0+F^2(u)du^2+R^2(u)(d\theta^2+\sin^2\theta d\p^2)~~\mathrm{at}~u<u_0,\nonumber\\
	&=& -\left[1-\frac{Q_m}{R(u)}\right]^2dt^2 + \frac{R^{'2}(u) du^2}{\left[1-\frac{Q_m}{R(u)}\right]^2} ~+ ~R^2(u)[d\theta^2+ \sin ^2\theta d\phi^2]~~\mathrm{at}~u>u_0
	\end{eqnarray}
	  	  The smooth and monotonic functions $F(u),R(u)$ and their $n$-th derivatives satisfy the following boundary conditions:
	\begin{eqnarray}\label{bc}
	&& F(u)\rightarrow 0, F^{(n)}(u)\rightarrow 0 \quad as \quad u\rightarrow u_0, \nonumber\\
			&&	R(u)\rightarrow Q_m ,~R^{(n)}(u)\rightarrow 0  \quad as\quad u\rightarrow u_0,\nonumber\\
			&& R(u)\rightarrow \infty,~ R'(u)=const. \quad as \quad  u\rightarrow \infty.
					\end{eqnarray}
					The function $R(u)$ may be interpreted as the common radius of the two-sphere almost everywhere.
							
We shall now show that the spacetime defined above reflects the presence of a (apparent) magnetic charge.
						
\subsection{A magnetic phase in Einstein gravity}
	 At $u>u_0$, the metric (\ref{metric}) is invertible ($\det g_{\mu\nu}\neq 0$) and becomes flat at the asymptotic boundary $u\rightarrow \infty$.	
	Under a local reparametrization of the coordinate $u\rightarrow R(u)$, this becomes equivalent to the extremal Reissner-Nordstrom metric at $u>u_0$. However, such an identification cannot hold everywhere, in particular, at the phase boundary $u=u_0$ where the metric smoothly degenerates. 
	
	The tetrad fields associated with this metric (the internal metric being $\eta_{IJ}=\left[-1,1,1,1\right]$) completely determine the torsionless spin-connection fields $\omega_{\mu}^{IJ}(e)$, which in turn lead to the following field-strength components:
	\begin{eqnarray}\label{R}
		R^{01}&=&\frac{Q_m}{R^3}\left( 2-\frac{3 Q_m}{R}\right)R' dt\wedge du ,\nonumber\\
		R^{02}&=&-\frac{Q_m}{R^2}\left(1-\frac{Q_m}{R}\right)^2 dt\wedge d\theta,\nonumber \\
		R^{03}&=&-\frac{Q_m}{R^2}\left(1-\frac{Q_m}{R}\right)^2\sin\theta dt\wedge d\phi,\nonumber\\ R^{12}&=&-\frac{Q_m}{R^2}R' du\wedge d\theta,\nonumber\\
		R^{23}&=&\frac{Q_m}{R}\left( 2-\frac{Q_m}{R}\right) \sin\theta d\theta \wedge d\phi,\nonumber\\ R^{31}&=&-\frac{Q_m}{R^2}R' \sin\theta d\phi\wedge du  .
		\end{eqnarray}
	  %Note that all the components vanish in the limit $u\rightarrow u_0$ except $R^{23}\goes \sin\theta d\theta\wedge d\phi$.

	In the matter sector, the only non vanishing component is the magnetic field:
	\begin{equation}
		F_{\theta\phi}=Q_m\sin\theta
	\end{equation}
This is characteristic of a magnetic pole of charge `$Q_m$'. 

The complete specification of the field content at the invertible phase as given above still does not tell us what exactly seeds the charge, or namely, whether it is genuine matter or geometry. The real origin of this monopole field depends on its continuation at $u\leq u_0$. This is what concerns us in the subsequent analysis.	
	
\subsection{Spacetime solution in noninvertible phase}							
At $u\leq u_0$, the tetrad (metric) fields in (\ref{metric}) have no inverse. These, along with an appropriate set of connection fields, define a degenerate spacetime solution of the first order Lagrangian density:
%, which do not require the invertibility of tetrad either in its definition or its variation :
\begin{eqnarray*}\label{L}
{\cal L}=\epsilon^{\mu\nu\alpha\beta}\epsilon_{IJKL} e_\mu^I e_\nu^J R_{\alpha\beta}^{~~KL}(\omega)
\end{eqnarray*}
The associated field equations in vacuum read \cite{tseytlin}:
\begin{eqnarray}\label{eom}
	\epsilon^{\mu\nu\alpha\beta}\epsilon_{IJKL} e_\nu^K D_{\alpha}(\omega)e_\beta^{L}&=&0,\nonumber\\ \epsilon^{\mu\nu\alpha\beta}\epsilon_{IJKL} e_\nu^J R_{\alpha\beta}^{~~KL}(\omega) &=&0
\end{eqnarray}
where the covariant derivative $D_\mu(\omega)$ is defined with respect to the connection $\omega_{\mu}^{IJ}$ as implied by its argument.

In general, any degenerate metric solution to the field equations (\ref{eom}) could exhibit nontrivial contortion. More precisely, the general solution for the nonvanishing components of the connection fields is given by \cite{kaul1}:
\begin{eqnarray}
\omega_a^{~ij}=\bar{\omega}_a^{~ij}(e)+\epsilon^{ijk} e_a^l N_{kl},~\omega_a^{~0i}=M^{il}e_{al},
\end{eqnarray} 
where the symmetric rank-three  matrices $N_{kl}$ and $M_{kl}$ encode contortion. Here, however, it is sufficient to assume $M_{kl}=0$, as these components play no role in the physics that concerns us here.
The resulting expression for the connection fields (with contortion) becomes:
	 \begin{eqnarray}\label{connections}
\omega^{12}&=& F \eta_2 du +\left(R\eta_1-\frac{R'}{ F}\right) d\T +\gamma H \sin\T d\p,\nonumber\\
		\omega^{23}&=& F \alpha  du + \eta_3 R d\theta +\left(\eta_2 R\sin\theta -\cos\theta \right)d\phi,\nonumber\\
		\omega^{31}&=& \eta_3 F du + \beta R d\theta +\left(R\eta_1+\frac{R'}{ F}\right)\sin\theta d\phi,	\nonumber\\
		\omega^{0i}&=&0,	
			\end{eqnarray}
where we use an explicit parametrization for the $N_{kl}$ field as:
\begin{eqnarray*}\label{N}
N_{ij}~=~\left(\begin{array}{ccc}
\alpha & \eta_3 & \eta_2\\
\eta_3 & \beta & \eta_1\\
\eta_2 & \eta_1 & \gamma\end{array}\right) 
\end{eqnarray*}
%%%%%%%%%%%%%%%
%%%%%%%%%%%%%%%%%
Let us now analyze the structure of the field-strength based on these connection fields:
\begin{eqnarray}\label{curv}
		&&R^{0i}=0\nonumber\\
		%%%%%%%%%%%%%%%%%%%%%%%%%%%%%%%%%%%%%%%% \left(\right)
		&&R^{12}=  \left[\frac{1}{FR}\left(\eta_1 R-\frac{R'}{ F}\right)'+(\eta_3^2-\alpha\beta)  \right]e^1\wedge e^2 \nonumber\\ &&+ \left[\frac{(\G-\B)}{R}\cot\T+\left(\B\e_2 -\e_3\left(\e_1 R+\frac{R'}{FR}\right)\right)\right]e^2\wedge e^3 \nonumber\\ & &+ \left[\frac{1}{FR}\left[-(\G R)'+\A R'\right]+(\A \e_1-\e_2\e_3)+ \frac{\eta_3}{R}\cot\T\right]e^3\wedge e^1\nonumber\\
		%%%%%%%%%%%%%%%%%%%%%%%%%%%%%%%%%%%%%%
		&&R^{23}= \left[\frac{(\e_3 R)'}{FR}+ \left(\e_2\B-\e_1\e_3+\frac{\eta_3 R'}{ FR}\right)\right] e^1\wedge e^2 \nonumber\\ && +\frac{1}{R^2}\left[\eta_2 R\cot\T +\left(1+(\e_1^2-\B\G)R^2-\left(\frac{R'}{F}\right)^2\right)\right]e^2\wedge e^3 \nonumber\\ & &+\left[-\frac{(\e_2 R)'}{FR}+ \left(\e_3\G -\e_2\left(\e_1 +\frac{H'}{FR}\right)\right)\right]e^3 \wedge e^1\nonumber\\
		%%%%%%%%%%%%%%%%%%%%%%%%%%%%%%%%%%%%%%%%%%
		&&R^{31}= \left[\frac{(\B R)'}{FR}+\left(\A\left(\eta_1 -\frac{R'}{ FR}\right)-\e_2\e_3 \right)\right]e^1\wedge e^2\nonumber\\ &&+ \frac{1}{R^2}\left[2\e_1 R\cot\T+\left(\left(\e_3\G-\e_1\e_2\right)R^2+\frac{R R'}{ F}\e_2\right)\right]e^2\wedge e^3\nonumber\\ & &+\left[-\frac{1}{FR}\left(\e_1 R+\frac{R'}{F}\right)'+\e_2
		^2-\A\G-\frac{\e_2}{R}\cot\T\right]e^3\wedge e^1 \nonumber\\
		\end{eqnarray}	
%%%%%%%%%%%%%%%
%%%%%%%%%%%%
An inspection of these reveals that the geometry exhibits a Dirac-string singularity along $\theta=0$ or $\theta=\pi$, unless some of the contortion fields vanish:
 \begin{eqnarray}\label{ehat}
 \eta_1=\eta_2=\eta_3=0,~\beta-\gamma=0.
 \end{eqnarray}
We shall impose this requirement here on. 

Given the fields above, all the field equations are satisfied except the component $\mu=t,I=0$ in the second set in eq.(\ref{eom}). In terms of the metric functions $F(u),R(u)$ introduced in (\ref{metric}), this reduces to:
	\begin{equation}\label{MC}
	2R\left[\frac{R'}{F}\right]'+\left[\left(\frac{R'}{ F}\right)^2-1\right]F+\beta[2\alpha+\beta] F R^2=0
	\end{equation}
	For any function $\Omega(u)\equiv \beta(2\alpha+\beta)(u)$, this equation has the general solution given below:
	\begin{eqnarray*}
F(u)= \frac{R'(u)}{\sqrt{1-\frac{1}{R(u)}\int du~\Omega(u) R^2(u)R'(u)+\frac{k}{R(u)}}}	
	\end{eqnarray*}
		where $k\geq 0$ is an integration constant.

	%In terms of the variable $\Phi(u)=\left[\left(\frac{R'}{\Ss F}\right)^2-1\right]$, we rewrite the above equation in a simpler form as:
	%\begin{eqnarray}
	%\left[R(u)\Phi(u)\right]'=3\lambda^2 R^2(u) R'(u).
	%\end{eqnarray}
	
	For simplicity, we shall consider the case where nontrivial contortion components $\alpha,\beta,\gamma$ are constant. 
	%The solution then reads:
	%	\begin{eqnarray}
	%F= \frac{R'}{\sqrt{1+\frac{1}{3}\left[2\beta(\alpha-\beta)Q_m^2-3\right]\frac{Q_m}{R}-\frac{1}{3}\left[\beta(2\alpha+\beta)\right] R^2}}
	%\end{eqnarray}
	With this, the solution for the four metric at the noninvertible phase finally becomes:
	\begin{eqnarray}\label{g1}
	ds^2=0+\frac{R^{'2}du^2}{\left[1-\frac{1}{3}\Omega R^2+\frac{k}{R}\right]}+R^2\left[d\T^2 +\sin^2 \T d\phi^2\right],
	\end{eqnarray}
	where the constants are given by $k=\frac{1}{3}\left[2\beta(\alpha-\beta)Q_m^2-3\right]Q_m, ~\Omega=\left[\beta(2\alpha+\beta)\right]$ and $k$ is fixed by the continuity of fields across $u=u_0$. 
Since $k>0$, the denominator in $g_{uu}$ has (at least) one positive real root (corresponding to the surface $R(u)=R_*(Q_m)$ for some $u=u_*>0$) in general. This defines the inner boundary of the metric.

\subsection{Smoothness of spacetime geometry}
The tetrad components defined by metric (\ref{metric}) are smooth by construction. 
Further, let us look at the torsion components at the zero determinant phase:
\begin{eqnarray}
T^0&=&0,\nonumber\\
T^1&=& -\beta R^2 \sin\theta~d\theta\wedge d\phi,\nonumber\\
T^2&=& -\frac{1}{2}(\alpha+\beta)F R \sin\theta~d\phi\wedge du,\nonumber\\
T^3&=& -\frac{1}{2}(\alpha+\beta)F R \sin\theta~du\wedge d\theta,
\end{eqnarray}
whereas the invertible phase has no torsion. 
Evidently, all components above smoothly go to zero as the phase boundary is approached, and are also finite at the inner boundary.
Finally, the $SO(3,1)$ field strength components in eq.(\ref{curv}), upon using (\ref{ehat}), read:
	\begin{eqnarray}\label{R}
						%%%%%%%%%%%%%%%%%%%%%%%%%%%%%%%%%%%%%%%% 
		R^{0i}&=&0,	\nonumber\\				
		R^{12}&=&  \left[-\left(\frac{R'}{F}\right)'-\alpha\beta FR \right]du\wedge d\theta\nonumber\\
		 &+&\left[\A R'-(\beta R)'\right]\sin\T d\p\wedge du,\nonumber\\
		%%%%%%%%%%%%%%%%%%%%%%%%%%%%%%%%%%%%%%
		R^{23}&=& \left[1-\beta^2 R^2-\left(\frac{R'}{F}\right)^2\right]\sin\T d\T\wedge d\p \nonumber\\
		%%%%%%%%%%%%%%%%%%%%%%%%%%%%%%%%%%%%%%%%%%
		R^{31}&=&\left[-\left(\frac{R'}{F}\right)'-\alpha \beta FR \right]\sin\T d\p\wedge du+ \left[(\beta R)'-\A R'\right] du\wedge d\T 
		\end{eqnarray}
%%%%%%%%
%%%%%%
A comparison with eq.(\ref{R}) shows that these are smooth as well. Since these fields ($e_\mu^I,~T_{\mu\nu}^{~~I}\equiv\frac{1}{2}D_{[\mu}e_{\nu]}^I,~R_{\mu\nu}^{~~IJ}$) define the complete set of gauge covariant fields in the first order formulation,  the full spacetime is manifestly smooth everywhere.

\section{Emergent magnetic field} 
The noninvertible phase given by (\ref{metric}) at $u\leq u_0$ naturally defines a set of (invertible) triad fields:
	\begin{eqnarray*}
	\hat{e}_u^1= F(u),~\hat{e}_\theta^2=R(u),~\hat{e}_\phi^3=R(u)\sin\theta,
	\end{eqnarray*} 
These, alongwith the unique torsionless connection $\hat{\omega}_a^{~ij}(\hat{e})=\bar{\omega}_a^{~ij}\equiv \frac{1}{2}[\hat{e}^b_i\del_{[a}\hat{e}_{b]}^j-\hat{e}^b_j\del_{[a}\hat{e}_{b]}^i-\hat{e}^a_l \hat{e}^b_i \hat{e}^c_j\del_{[b} \hat{e}_{c]}^l]$, describe an effective Einstein theory in three space. The contortion $K_a^{~ij}$, however, is not a part of this geometry. Rather, it lives as a (emergent) matter field on it. The precise nature of such matter must emerge naturally from the basic equations within this formulation. 

In view of the fact that all the field equations (\ref{eom}) have already been solved, let us consider the identity involving torsion and the hodge dual of the $SO(3,1)$ field-strength:
\begin{eqnarray}
\epsilon^{\alpha\beta\rho\sigma}D_\beta(\omega) T_{\rho\sigma}^{~~I}=\frac{1}{2}\epsilon^{\alpha\beta\rho\sigma}
R_{\beta\rho}^{~~IJ}(\omega)e_{\sigma J}
\end{eqnarray}
The $\alpha=t,~I=i$ component of the above, upon projection along a purely spatial unit vector $n^i$ in the internal space, leads to:
\begin{eqnarray}
\del_a\left[\frac{1}{2}\epsilon^{abc}T_{bc}^{~~i}n_i\right]=\frac{1}{4}\epsilon^{abc}R_{ab}^{~~ij}n_i e_{cj}+
\frac{1}{2}\epsilon^{abc}T_{bc}^{~~i}D_a n_i\nonumber\\
\end{eqnarray}
Given the even parity of torsion, this equation is precisely of the form $\del_a(\hat{e}\hat{B}^a)=\hat{e}\hat{\rho}_m$ in curved space, where the (static) magnetic field is sourced by a nontrivial magnetic charge density, defined as:
\begin{eqnarray}\label{B}
&&\hat{e}\hat{B}^a\equiv -\frac{1}{2}\epsilon^{abc}T_{bc}^{~~i}n_i\equiv \frac{1}{2}\epsilon^{abc}F_{bc},\nonumber\\
&&\hat{e}\hat{\rho}_m\equiv \frac{1}{4}\epsilon^{abc}R_{ab}^{~~ij}n_i e_{cj}+
\frac{1}{2}\epsilon^{abc}T_{bc}^{~~i}D_a n_i
\end{eqnarray}
To emphasize, the vector $n^i$ defines the projection from the $SO(3)$ fields to the (emergent) $U(1)$ field-strength in this context.

A radial field may now be obtained by choosing $n^i$ to point radially: $n^i=(1,0,0)$.
Inserting the solution at the degenerate phase constructed explicitly earlier, we find:
\begin{eqnarray}
\hat{F}_{\theta\phi}=\beta R^2\sin\theta,
\end{eqnarray}
Note that this is independent of $\alpha$. Remarkably, it approaches a monopole field ($\hat{F}_{\theta\phi}\rightarrow Q \sin\theta$) at the outer boundary, and is nonsingular everywhere all the way upto the inner boundary. The continuity of the field-strength across $u=u_0$ fixes the free constant above as: $\beta=\frac{1}{Q_m}$.

To see if the emergent definition (\ref{B}) does reproduce the correct physics in this context, let us consider the integral of the charge density at the phase boundary:
	\begin{eqnarray}
\frac{1}{4\pi}\int d^3 x~\hat{e}\hat{\rho}_m&=&\frac{1}{4\pi}\int_{S^2} du~ d\theta~d\phi~ \del_u\left[\beta R^2\sin\theta\right] =Q_m\nonumber\\
&=&\frac{1}{4\pi}\int_{S^2} d\T d\phi~ \hat{F}_{\T\phi},
	\end{eqnarray}
	where in obtaining the first equality above we have used the following identities:
	\begin{eqnarray*}
	&&\frac{1}{4}\epsilon^{abc}R_{ab}^{~~ij}n_i \hat{e}_{cj}=(\beta-\alpha)RR'\sin\theta,\\
	&&\frac{1}{2}\epsilon^{abc}T_{bc}^{~~i}D_a n_i= (\alpha+\beta)RR'\sin\theta~.
	\end{eqnarray*}
 %This core has a finite size $\sim d(\lambda,Q)=R(u_0)-R(u_*)$, determined by the scale parameter $\lambda$ and the charge $Q$.	
	Thus, the net flux through the (compact) phase-boundary is indeed equal to the magnetic charge of the solution.

	This completes the demonstration how a static nonsingular monopole could emerge as an artefact of the noninvertible vacuum phase of gravity theory in general. Even though we have considered the simpler case of constant contortion, a generalization to the inhomogeneous case is straightforward. However, that does not add to the insights gained here, since the essential physics depends on their value at the phase boundary, which remain unaffected.

%%%%%%%%%%%%%%%%%%%%	
%%%%%%%%%%%%%%%	
%	Since the scalars built out of the three curvature tensors $ \tilde{R}_{ab}^{ij} $ are all finite everywhere, the non-degenerate 3-geometry at $u\leq u_0$ does not exhibit an curvature singularity in this sense.. This  feature should be contrasted to the case of the standard continuation of the RN external solution at $r\leq Q$, which corresponds to a curvature singularity in the no-degenerate 4-geometry. This also suggests a possibility way out of the issue of the point charge singularity in Maxwell's theory. Our analysis shows that in first  order gravity where the inevitability of metric is not essential, such singularities need not arise in general.
	
	\section{A topological origin of apparent magnetic charge}
%In this formulation, the apparent magnetic charge has a topological interpretation in terms of a geometric invariant that could be constructed at the zero-determinant phase. 
Now we explore if it is possible to construct a geometric invariant, which could encode the boundary (topological) effects induced by torsion at the surface separating the two phases. The boundary here has the topology of $S^2$. The only torsional invariant that may be defined on it is given by:
\begin{eqnarray}\label{N}
N&=&-\frac{1}{8\pi} \int_{S^2_B} d^2 x ~\epsilon^{ab}T_{ab}^{~~i}n^i
\end{eqnarray}
where $a\equiv(\theta,\phi)$.
This is simply the two-integral of a unique gauge-invariant projection of the torsion two-form over a compact manifold. This, however, may be seen as a lower dimensional counterpart of the Nieh-Yan invariant, which is a four-integral (over a compact manifold) of a gauge-invariant projection of the appropriate torsional four-form \cite{nieh}.
%:
%\begin{eqnarray*}
%N_{NY}=-\frac{1}{16\pi^2}\int d^4 x~ \partial_\alpha \left[\epsilon^{\alpha\beta\rho\sigma} T_{\beta\rho}^{~~I}e_{\sigma I}\right]
%\end{eqnarray*}
%We are interested 
%	Let us consider the Chern-Simons current associated with the $SO(3,1)$ connection $\hat{\omega}_a^{~IJ}\equiv(\hat{\omega}_a^{~ij},~\hat{\omega}_a^{~4i}=\lambda\hat{e}_a^{i})$:
%	\begin{eqnarray}
%	\frac{1}{2}\epsilon^{tabc}\hat{\omega}^{IJ}_a\left[ \del_b \hat{\omega}_{cIJ}+\frac{2}{3}\hat{\omega}^{IK}_b \hat{\omega}_{cK}^{~~J}\right]=\frac{1}{2}\epsilon^{tabc}\hat{\omega}^{ij}_a\left[ \del_b \hat{\omega}_{cij}+\frac{2}{3}\hat{\omega}^{ik}_b\hat{\omega}_{ck}^{~~j}\right]-\lambda^2 \epsilon^{tabc}\hat{e}^i_a D_b(\hat{\omega}) \hat{e}_{ci}
	~
	%\nonumber\\&&=-2\lambda^2 \epsilon^{tabc}\hat{e}^i_a D_b(\hat{\omega}) \hat{e}_{ci}
%	\end{eqnarray}
%	Thus, we can construct a geometric invariant  by taking the difference of the Chern-Simons currents corresponding to the $SO(3,1)$ and $SO(3)$ connections $\hat{\omega}_a^{~IJ}$ and $\hat{\omega}_a^{~ij}$ respectively, and integrating it over the compact boundary hypersurface(s). These boundaries here are given by the two spheres ($S^2_B$), as  $u\rightarrow u_*$ and $u\rightarrow u_0$, respectively. Explicitly, this geometric invariant may be viewed as the lower dimensional counterpart of the Nieh-Yan index \cite{nieh}:
%	\begin{eqnarray}
%		N=-\frac{1}{8\pi} \int_{S^2_B} ds_a n^a \epsilon^{bcd}\hat{e}^i_b D_c(\hat{\omega}) \hat{e}_{di},
%	\end{eqnarray}
%	where $n^a \equiv n^i e^{a}_i = (\frac{1}{\sqrt{\hat{g}_{uu}}},0,0)$ is the unit normal on the two-sphere. 
	
Next, let us evaluate the reduced Nieh-Yan charge (\ref{N}) for the (monopole) geometry constructed earlier: 
\begin{eqnarray}\label{N1}
N=\frac{1}{4\pi}\int_{S^2_B} d\T d\phi~ \left[K_\theta^{~31}e_\phi^3+K_\phi^{~12}e_\theta^2\right]= Q_m,
\end{eqnarray}
where we have used the expressions $K_\theta^{~31}=\beta R,~K_\phi^{~12}=\gamma R\sin\theta=\beta R\sin\theta$ from (\ref{connections}).
As remarkable as it seems, the topological Nieh-Yan number at the noninvertible phase $u<u_0$ is precisely what gets reflected as the magnetic charge to an observer at the Einsteinian phase at $u>u_0$. The magnetic charge thus acquires a topological interpretation within a first order formulation of gravity theory in absence of matter.  

The associated winding number corresponds to a map from the $S^2$ in the coordinate space to the  $S^2$  in the internal space. This may be defined through a unit vector $m^i (\theta,\phi)\equiv (\sin\theta \cos\phi,\sin\theta \sin\phi,\cos\theta)$:
\begin{eqnarray*}
w=\frac{1}{8\pi}\int_{S^2_B}d^2 x~\epsilon_{ijk}\epsilon^{ab}m^i \del_a m^j \del_b m^k=1.
\end{eqnarray*}
%where the integral is evaluated at the phase boundary.
Note that in terms of $m^i$, the emergent three-metric at the noninvertible phase could be given by the following set of vielbein fields:
\begin{eqnarray*}
E^i=R\del_a m^i dx^a~[a\equiv \theta,\phi],~E^4=\sqrt{\sigma}Fdu.
\end{eqnarray*}

%The factor of $2$ reflects the fact that there are two independent fields on $S^2$ which contribute to the integral.
%\footnote{Strictly, the integral at the outer boundary has a regularization implicit in it.  Since the differential $dR$ fails to be analytic at the phase boundary $u=u_0$, 
%it could be made well-defined by introducing a small parameter $\epsilon$. 
% the integral should be evaluated an infinitesimal distance $\epsilon$ away from (inside) the boundary, before taking the regulator to zero.}.

\section{Conclusions}
%The seminal works of Dirac and t'hooft-Polyakov on gauge theory monopoles have left a long trail.
 From a modern perspective, the unobservability of free magnetic poles sourced by matter fields could be turned around to the assertion that such objects do not exist in nature, and their only manifestation could be emergent. Here we explicitly proposed such a realization
in curved spacetime with defects associated with a vanishing metric determinant\footnote{The reader may see ref.\cite{hanson} for one of the earliest discussions of torsion defects in a context different from this work. Such configurations, however, are not solutions to the vacuum field equations of gravity.}. This description of an emergent magnetic pole is based on a special class of classical solutions to the first order field equations of vacuum gravity, where the temporal direction exhibits a zero eigenvalue. 
Our construction has no analogue in standard Einsteinian gravity in vacuum based on invertible tetrad fields everywhere, since such a theory does not admit a classical solution with nontrivial torsion as is necessarily the case here.

In addition, we provide a topological interpretation of the (apparent) magnetic charge in gravity theory, given by a lower dimensional counterpart of the torsional Nieh-Yan topological number. The existence of a compact surface at the phase boundary makes such an identification possible.
 
The details as to how the apparent magnetic charge gets manifested to the observer necessarily living in an Einstein (invertible) phase have been elucidated. Based on smoothness requirements, this geometry is found to be the magnetic Reissner-Nordstrom solution (extremal).  
 Since there are no analytic radial geodesic from this outer region to the noninvertible phase, the monopole defect remains hidden. However, the detection of the topological charge of the defect is still possible through the gravitational potential away from the phase boundary.
 
The emergent field and also the three-geometry seeding the topological charge are both regular. This may be contrasted with the point curvature singularity characteristic of a magnetic Reissner-Nordstrom black hole. However, any statement regarding a possible
resolution of the curvature singularity in a four dimensional sense needs to
be formulated in terms of relevant fields superceding the four dimensional
curvature invariants which cease to be well defined themselves due the
noninvertibility of tetrad. It is an interesting question as to whether this regulating effect of the noninvertible phase could play any role in the renormalizability of quantum gravity in first order formulation where torsion is expected to introduce a scale. 

A relevant criticism of our construction could be that it is based on spacetimes with noninvertible metric phases, which have no place in standard Einstein theory. However, it would not be logical to ignore them based on this ground only, since these could mediate topology change in classical theory, a possibility that need not be excluded in principle.
As we have demonstrated here, some of these are special and could carry an apparent magnetic charge. It would certainly be important to understand the significance of these geometries beyond the classical realm. For instance, the inclusion of such `charged' saddle points in the functional integral could lead to fresh insights regarding the role of Nieh-Yan topological charge in quantum gravity.

Furthermore, when the timelike direction lies along the zero eigenvalue of the degenerate four-metric, one might as well be bothered by the apparent loss of causality at this phase. However, similar scenarios are typical in quantum field theory, which is known to admit virtual particles. Momenta of such
particles need not lie on the relativistic mass shell, and these need not respect causality all along their respective trajectories. In other words, the deeper implications of the apparent acausality \cite{sengupta1} associated with the noninvertible metric phase, once completely understood, could have their own virtue.

 %One could explore if this leads to a quantization of geometry in gravity theory from a purely classical reasoning. It is also worth noting that our formulation provides a natural connect between the emergent electromagnetism in curved spacetime and viscoelastic electromagnetism in the condensed matter context \cite{hidaka}. Systems such as these could be interesting  testbeds for some of the ideas presented here.

\begin{acknowledgments}
%\acknowledgments
%\acknowledgments 
The work of S.S. is supported (in part) by the ISIRD project grant `RAQ'. S.G. gratefully acknowledges the support of a DST Inspire Fellowship.
\end{acknowledgments}

\end{document}